\documentclass[usenatbib]{mn2e}
\usepackage{times}
\usepackage{epsfig}
\usepackage{amsmath}
\title[Stars and hot gas in low-mass galaxy clusters]{The stellar and hot gas content of low-mass galaxy clusters}
\author[Balogh et al.]{Michael L. Balogh$^{1}$, Pasquale
  Mazzotta$^{2,3}$, Richard G. Bower$^{4}$, Vince Eke$^{4}$,  
\newauthor Herv{\' e} Bourdin$^{2}$, Ting Lu$^{1}$, Tom Theuns$^{4,5}$
\\
$^{1}$Department of Physics and Astronomy, University of Waterloo, Waterloo, Ontario, N2L 3G1, Canada\\
$^{2}$Dipartimento di Fisica, Universit{\` a} degli Studi di Roma ``Tor Vergata'', via della Ricerca Scientifica, 1, 00133 Roma, Italy\\
$^{3}$Harvard-Smithsonian Center for Astrophysics, 60 Garden Street, Cambridge, MA 02138, USA\\
$^{4}$Department of Physics, University of Durham, Science Laboratories, South Road, Durham DH13LE, UK \\
$^{5}$Department of Physics, University of Antwerp, Campus Groenenborger, Groenenborgerlaan171, B-2020 Antwerp, Belgium\\
}
\date{\today}

\def\gtrsim{\mathrel{\raise0.35ex\hbox{$\scriptstyle >$}\kern-0.6em
\lower0.40ex\hbox{{$\scriptstyle \sim$}}}}
\def\lesssim{\mathrel{\raise0.35ex\hbox{$\scriptstyle <$}\kern-0.6em
\lower0.40ex\hbox{{$\scriptstyle \sim$}}}}

\def\k04{{^{0.4}K_s}}
\def\s04{{^{0.4}S1}}

\begin{document} 
\maketitle
\begin{abstract}
We analyse the stellar and hot gas content of 18 nearby, low-mass galaxy clusters,
detected in redshift space and selected to have a dynamical mass
$3\times10^{14}<M/M_\odot<6\times10^{14}$ ($h=0.7$), 
as measured from the  2dF Galaxy Redshift Survey.  We combine X-ray
measurements from both {\it Chandra} and {\it XMM} with ground-based
near-infrared observations from CTIO, AAT and CFHT to compare the mass
in hot gas and stars to the dynamical mass and state of the clusters.
Only 13 of the clusters are detected in X--ray emission, and for these
systems we find that a range of 7--20 per cent of their baryonic mass,
and $<3$ per cent of their dynamical mass, is detected
in starlight, similar to what is
observed in more massive clusters.   
In contrast, the five undetected clusters are underluminous in X--ray
emission, by up to a factor $10$, given their stellar mass.
Although the velocity
distribution of cluster members in these systems is indistinguishable
from a Gaussian, all show subtle signs of being
unrelaxed: either they lack a central, dominant galaxy, or the
bright galaxy distribution is  less concentrated and/or more elongated
than the rest of the sample.  Thus we conclude that low--mass clusters
and groups selected from the velocity distribution of their galaxies
exhibit a dichotomy in
their hot gas properties.  Either they are detected in X-ray, in which
case they generally lie on the usual scaling relations, or they are
completely undetected in X-ray emission.   
The non-detections may be partly related
to the apparently young dynamical state of the clusters, but it remains a
distinct possibility that some of these systems are
exceptionally devoid of hot emitting gas as the result of its expulsion or rarefaction.
\end{abstract}

\section{Introduction}\label{sec-intro}
Galaxy clusters and groups are important both as cosmological probes,
and as laboratories for studying galaxy evolution.  In particular, their
deep gravitational potential means that their baryon content should be nearly
representative of the Universe as a whole, and that the diffuse gas is at a
temperature that is accessible to observation.  As a result, they
represent one of the few places where it is possible to study the
stars, cold and hot gas, and dark matter in a single system.

It is now well known that the mass fraction in stars is not universal,
but in general decreases with increasing cluster mass
\citep[e.g.][]{Eke-groups2,Eke-groups3,RBGMR,Lin03,G+09}.  On the other hand, the mass fraction of
hot gas appears to {\it increase} with mass
\citep[e.g.][]{Vik+05,Sun08,Pratt09}.  If these systems are closed boxes,
then the sum of stellar and gas mass fractions should equal the universal value
$f_b=\Omega_b/\Omega_m$, where $\Omega_b$ and $\Omega_m$ are the baryon
and matter densities, respectively,  relative to the critical density.  Recently \citet{GZZ} claimed this to be the
case, in particular arguing that in the lowest mass systems there is
significant stellar mass in the intracluster-light and halo of the
central galaxy, which completes the baryon fraction.  The
interpretation then is that most of the baryons in these low-mass groups
have cooled to form stars; this poses a challenge for normal
hierarchical models which predict that more massive clusters are
actually built from these groups \citep{BMBE}.  An important open question then is whether or not there really exists a
significant population of groups with $M_{\rm stars}/M_{\rm gas}>20$
per cent.  The conclusions of
\citet{GZZ} depend partly on an extrapolated mean relation for the gas
fraction of clusters as a function of mass, taken from \citet{Vik+05}.
However, similar conclusions were reached by \citet{LLAC}, based on a
small sample of five nearby Abell clusters observed with {\it XMM}.

The other possible explanation for the high ratio of stellar-to-gas
mass in groups is that these systems are deficient in X--ray emitting gas, and associated metals. Recent observations by \citet{G+09} and \citet{RP09} indicate
this is the case; in particular the latter provides an interesting analysis of the abundances in a sample of 15
X--ray bright groups \citep{RP07} and concludes that these groups have
fewer metals than expected given their stellar mass.  However, their sample was selected
\citep[from the larger sample of][]{OP04} to have relaxed X--ray
morphologies and good photon statistics.  They may therefore be biased toward
undisturbed, cool--core groups which are unlikely to be the typical
precursor of more massive clusters \citep{RP09}.
The apparent requirement for strong energy sources to counter the high
cooling rates near the centre of clusters may provide a natural
mechanism for removing hot gas from groups.  Models incorporating supernova--driven superwinds \citep[e.g.][]{DOS08} or supermassive black hole accretion \citep[e.g.][]{Bower+08,McCarthy+10} have had considerable success at matching the observed properties of X--ray emitting gas in groups and clusters.  Alternatively, models in which gas was heated prior to the virialization of the group or cluster predict that such systems would fail to accrete their full complement of gas in the first place \citep[e.g.][]{entropy,holistic}.

Most of the work that has been done so far has been based on X--ray
selected samples of clusters, which introduces a potential
bias.  Several investigations have been conducted in an attempt to
evaluate the importance of this bias.  In particular,
\citet{BBB+94}, \citet{Gilbank04}, \citet{Donahue+01},
\citet{Popesso-V} and \citet{Rozo+08} make use of pointed and archival
{\it ROSAT} observations of cluster samples, identified based upon their
optical galaxy population using redshift and position information.  All of
these independent 
studies found significant scatter between the optical richness (or
total optical luminosity) and X--ray luminosity of their systems.  In
particular, both \citet{BCECB} and \citet{Rykoff1} find their optically-detected clusters are
systematically underluminous in X--rays, relative to X--ray selected
samples.  Similar conclusions have been reached by
\citet{Hicks08}, using {\it Chandra} observations of 13
optically--selected clusters at $0.6<z<1.1$.   This is at least partly
due to a Malmquist-type bias in X--ray flux limited surveys
\citep{Ikebe,PCAB,Vik+09}; for example, \citet{Rykoff2} use weak-lensing based mass estimates to conclude that,
once a correction is made for non-hydrostatic equilibrium and this 
Malmquist bias, the $L_X-M$ relation defined by their
optically--selected clusters is similar to that of the X--ray selected
HIFLUGCS \citep{HIFLUGCS} sample.  

Recently, \citet{XI-I}
have begun an interesting study of a redshift--selected, unbiased group
sample followed up with {\it XMM}.  Of the first nine analysed, only
three were detected, and two of these are underluminous given their velocity
dispersion \citep{Bai+10}.  Based on their velocity distribution, \citet{XI-I} argue that the underluminous
systems are in the early stage of collapse; similar interpretations of such systems have been made by \citet{Popesso-V} and \citet{D+09}.

\begin{table*}
\begin{tabular}{lllllllll}
Id&Name&RA&Dec&Redshift&NIR source&Image size\\
  &      &\multispan2{\hfil(J2000)\hfil}& & &(arcmin)\\
\hline
1&  A2734        &   2.83863& -28.80177&  0.061& AAT  & $21.7\times 21.7$\\
2&  A3880        & 336.97528& -30.57532&  0.058& AAT  & $21.7\times 21.6$\\
3&  A3094        &  47.90881& -26.85087&  0.068& 2MASS& $39.0\times 55.5$\\
4&  A1650        & 194.69207&  -1.80407&  0.084& CFHT & $27.4\times 27.4$\\
5&  RBS317       &  36.294  & -29.486  &  0.060& AAT  & $21.7\times 21.7$\\
6&  MS1306.7-0121& 197.31057&  -1.61012&  0.086& CTIO & $31.8\times 31.8$\\
7&  S0041        &   6.38084& -33.04638&  0.050& 2MASS& $38.8\times 65.5$\\
8&  A954         & 153.43704&  -0.12043&  0.095& CFHT & $27.3\times 27.4$\\
9&  A1663        & 195.60600&  -2.52938&  0.083& 2MASS& $39.6\times 65.4$\\
10& Chan4990     & 198.03166&  -0.98243&  0.084& CFHT & $27.3\times 27.4$\\
11& XMM5        &  48.80986& -29.14884&  0.068& 2MASS& $38.6\times 55.5$\\
12& Chan4991     & 170.80439&   1.08753&  0.074& CTIO & $31.7\times 31.9$\\
13& XMM3         &   8.96560& -27.52812&  0.071& 2MASS& $38.8\times 55.4$\\
14& XMM9         & 201.76353&   1.31428&  0.081& CFHT & $27.3\times 27.4$\\
15& XMM4         & 199.83071&  -0.88001&  0.084& 2MASS& $30.0\times 30.0$\\
16&              & 150.93404&  -2.17231&  0.096& CTIO & $31.8\times 31.9$\\
17& XMM10        & 202.61707&   1.35255&  0.082& 2MASS& $47.1\times 52.0$\\
18&              & 349.50853& -28.17449&  0.077& AAT  & $21.6\times 21.7$\\
\hline
\end{tabular}
\caption{Basic coordinates of all clusters in our sample, including the source of the infrared imaging used in this paper.  \label{tab-groups}}
\end{table*}
\begin{table*}
\begin{tabular}{lllllllll}
Id&$\sigma$&$M_{\rm dyn}$     &$R_{\rm rms}$&$M_{\rm 200}$     &$R_{\rm 200}$&$M_{\rm 500}$     &$R_{\rm 500}$&$r_{\rm max}$\\
  &(km/s)  &($10^{14}M_\odot$)&(Mpc)        &($10^{14}M_\odot$)&(Mpc)        &($10^{14}M_\odot$)&(Mpc)&($R_{500}$)\\
\hline
1&590$\pm$54&3.6$\pm$0.8&0.89$\pm$0.04&3.25$\pm$0.82&1.34$\pm$0.12&7.06$\pm$2.27&1.32$\pm$0.14&0.64\\
2&580$\pm$58&4.8$\pm$1.0&1.23$\pm$0.04&3.10$\pm$0.69&1.32$\pm$0.13&3.09$\pm$1.17&1.00$\pm$0.12&0.80\\
3&700$\pm$53&5.0$\pm$1.0&0.88$\pm$0.04&5.37$\pm$1.25&1.57$\pm$0.12&1.49$\pm$0.42&0.78$\pm$0.07&1.4\\
4&600$\pm$54&5.6$\pm$1.2&1.34$\pm$0.04&3.31$\pm$0.81&1.32$\pm$0.12&6.94$\pm$0.17&1.30$\pm$0.01&0.71\\
5&550$\pm$64&3.5$\pm$1.0&1.00$\pm$0.04&2.63$\pm$0.91&1.25$\pm$0.15&1.24$\pm$0.10&0.74$\pm$0.02&0.80\\
6&540$\pm$69&3.8$\pm$1.2&1.12$\pm$0.07&2.40$\pm$0.82&1.18$\pm$0.15&1.40$\pm$0.09&0.76$\pm$0.02&0.81\\
7&580$\pm$49&4.1$\pm$0.9&1.05$\pm$0.04&3.13$\pm$0.81&1.33$\pm$0.11&6.04$\pm$3.37&1.21$\pm$0.27&0.26\\
8&680$\pm$72&4.5$\pm$1.1&0.84$\pm$0.06&4.74$\pm$1.26&1.47$\pm$0.16&1.00$\pm$0.14&0.68$\pm$0.03&0.70\\
9&620$\pm$63&5.0$\pm$1.1&1.12$\pm$0.04&3.65$\pm$0.96&1.36$\pm$0.14&1.68$\pm$0.02&0.81$\pm$0.00&0.94\\
10&530$\pm$70&3.1$\pm$0.9&0.95$\pm$0.05&2.28$\pm$0.72&1.16$\pm$0.15&2.12$\pm$0.88&0.87$\pm$0.12&1.1\\
11$^{*}$&570$\pm$59&3.3$\pm$0.8&0.87$\pm$0.05&2.90$\pm$0.85&1.28$\pm$0.13&$<0.18$&$<0.38$&\\
120&560$\pm$57&3.4$\pm$0.8&0.93$\pm$0.04&2.73$\pm$0.77&1.24$\pm$0.13&0.91$\pm$0.45&0.66$\pm$0.10&0.66\\
13$^{*}$&640$\pm$59&3.6$\pm$1.1&0.76$\pm$0.08&4.08$\pm$1.27&1.43$\pm$0.13&$<0.18$&$<0.38$&\\
14&720$\pm$61&5.5$\pm$1.2&0.91$\pm$0.06&5.74$\pm$1.32&1.59$\pm$0.13&0.55$\pm$0.07&0.56$\pm$0.02&0.80\\
15&700$\pm$69&5.6$\pm$1.3&0.98$\pm$0.05&5.25$\pm$1.39&1.53$\pm$0.15&0.74$\pm$0.37&0.61$\pm$0.08&0.54\\
16$^{*}$&480$\pm$111&3.2$\pm$1.3&1.19$\pm$0.09&1.66$\pm$0.77&1.04$\pm$0.24&$<0.18$&$<0.38$&\\
17$^{*}$&730$\pm$97&4.3$\pm$1.7&0.69$\pm$0.09&5.97$\pm$2.59&1.60$\pm$0.21&$<0.18$&$<0.38$&\\
18$^{*}$&710$\pm$94&5.2$\pm$2.1&0.89$\pm$0.07&5.53$\pm$2.59&1.57$\pm$0.21&$<0.18$&$<0.38$&\\
\hline
\end{tabular}
\caption{Dynamical properties, mass and radii measurements for each cluster used in this paper. The radii $R_{\rm rms}$ and $R_{\rm 200}$, with their associated mass estimates  $M_{\rm dyn}$ and $M_{200}$ are computed from the galaxy redshifts and positions as described in the text.  $M_{500}$ and
$R_{500}$ are measured from the X--ray images, and $r_{\rm max}$ is
the radius within which X-ray emission is detected, in units of $r_{500}$.  
The starred entries are undetected in X-ray, and for these systems we
calculate upper limits on $R_{500}$ and $M_{500}$ by extrapolating the
correlation between these quantities and $L_{\rm bol}$.  Groups 16 and
18 do not have {\it Chandra} or {\it XMM} observations.  \label{tab-groups_dyn}}
\end{table*}

What is clear is that the mass and dynamical state distribution of a
cluster sample will be sensitive to the way the clusters are detected.  It is
important to understand these effects, both to identify robust
indicators of a cluster's mass and to use them as cosmological probes.
Perhaps even more interesting is the fact that the scatter in key
properties such as gas entropy and stellar fraction, at fixed total
mass, contains important information about the physical processes
associated with galaxy formation \citep[e.g.][]{OWLS1,DOS}.  

One of the
most common ways to select groups and clusters from large redshift
surveys is the friends-of-friends linking method, which has the
advantage that it is easy to implement and
fairly straightforward to calibrate with numerical simulations
\citep[e.g.][]{Eke-groups}.  \citet{Eke-groups3} used such a sample
from the 2dFGRS \citep{2dF_colless} to demonstrate that most of the
stars in the Universe are associated with small groups,
$M\sim2\times10^{12}M_\odot$, in reasonable agreement with model
predictions \citep{Bower05}.   It would be of interest to know
the fate of the hot gas in groups selected in this way: in particular,
what is the efficiency of converting gas to stars on average, and what
is its variation between systems?

We have therefore selected 18 low-mass clusters, with
$3\times10^{14}<M/M_\odot<6\times10^{14}$,
from the catalogue of
\citet{Eke-groups}.   All but two of the 18 clusters were
followed up with pointed {\it Chandra} or {\it XMM} observations.  
This represents the first time that a complete, optically--selected sample of
low-mass clusters has been followed up with these facilities.
Surprisingly, we found that despite the narrow selection on dynamical
mass, the clusters have a wide range of X-ray luminosities, and five of them were
undetected.  The
details of the object selection and X--ray observations are given in
Mazzotta et al. (in prep, hereafter Paper~II).  Here we present
an analysis of near-infrared (NIR) data in these clusters, to measure
the stellar luminosity and mass.  The plan of the paper is as follows.
The origin and reduction of the NIR data is described in
\S~\ref{sec-obs}.  The details of the photometry and calculation of
NIR luminosity are given in \S~\ref{sec-analysis}.  
We then calculate the total NIR luminosity of each
cluster and consider various scaling relationships with dynamical and
X--ray properties, in \S~\ref{sec-results}.  We find that most of the clusters appear normal when compared with X--ray selected samples; we explore possible explanations for the five undetected clusters in 
\S~\ref{sec-discuss}, finally drawing our conclusions in
\S~\ref{sec-conc}.

We use a cosmology with $\Omega_m=0.3$, $\Omega_\Lambda=0.7$ and
$h=H_\circ/(100$km/s/Mpc)$=0.7$.  All NIR magnitudes are on the Vega (2MASS) system.

\section{Observations}\label{sec-obs}
\subsection{Cluster selection}
We selected 18 clusters from the catalogue of \citet{Eke-groups}; the
adopted centres and redshifts, which in some cases differ from those in
the original catalogue, are given in Table~\ref{tab-groups}.  Our main aim was to look for the variation in
X--ray emission from a mass-selected sample; thus we considered all
groups from that catalogue with 
$3\times10^{14}<M/M_\odot<6\times10^{14}$.   Our default definition of the mass, radius and velocity dispersion come
from \citet{Eke-groups}, where these quantities are related by
\begin{equation}
M_{\rm dyn}=\frac{5}{G}R_{\rm rms}\sigma^2.
\end{equation}
We excluded a few groups from this selection with large uncertainties
on the dynamical mass, due to poor membership or clearly non-Gaussian
velocity dispersions.  All selected groups have redshifts for at least
15 members.

The velocity dispersions are computed using the gapper estimate of
\citet{Beers}, and $R_{\rm rms}$ is the weighted {\it rms} projected
separation from the cluster centre of all members.  The factor of 5 was chosen to give a mass which is in good agreement with dark
matter haloes identified in numerical simulations, using a
friends-of-friends algorithm with linking parameter $b=0.2$.  Therefore
the mass $M_{\rm dyn}$ corresponds to the mass on scales larger than
$R_{\rm rms}$.
We recalculate 
uncertainties on $\sigma$ and
$R_{\rm rms}$ using a jackknife technique.
Uncertainties on $\sigma$ are typically 10--20 per cent, while those on
the radius are typically 5--10\%.  These lead to dynamical mass
estimates which are uncertain by 20--40\%.

For comparison with the literature it will also be useful to define a dynamical estimate of the "virial radius", $R_{\rm 200}$.  Following  \citet{RBGMR} we define
\begin{equation}\label{eqn-r200}
R_{\rm 200}=\frac{\sqrt{3}\sigma}{10H_\circ(1+z)^{-3/2}}
\end{equation}
and 
\begin{equation}\label{eqn-m200}
M_{\rm 200}=\frac{3}{G}R_{\rm 200}\sigma^2.
\end{equation}
In contrast with $M_{\rm dyn}$, this mass depends only on the measured
velocity dispersion, and is approximately independent of the radial extent of the
friends-of-friends group, since in general $\sigma$ is a weak function
of radius outside the core.  

The original 2dFGRS imposed a bright magnitude
  limit on the spectroscopic selection \citep{2dF_colless}.  As a result, the brightest
  galaxies in several of our clusters do not have a redshift from the
  original survey.  Many of these were found in the NASA Extragalactic Database and in the 6dFGRS survey
  \citep{6dF}.  In three clusters (8, 10, 11) there are still fairly
  bright galaxies without redshifts,
  but they are subdominant and it makes little difference whether we
  include them as cluster members or not (we do).  For cluster 13, the
  central very bright galaxy has no redshift available, and this
  contributes substantially to the cluster luminosity.  As it is
  centrally located we feel confident that it is a cluster member, and
  thus throughout the paper we assume this is the case.

\subsection{X-ray Observations}\label{sec-xray}
The description of the acquisition,
reduction and analysis of X-ray data from {\it Chandra} and {\it XMM}
are given in Paper~II.   Here we summarize the salient details.
{\it Chandra} data have been analyzed using {\sc CIAO} v4.0 and  {\sc CALDB} v3.4.3, and the standard tools. 
{\it XMM} data have been analyzed using the procedure described in \citet{BM08}.
For each cluster we extract the surface brightness and temperature
profiles which are then used 
to estimate the gas and total mass profile following 
the approach proposed by \citet{Vik+05}.  This procedure, which assumes
hydrostatic equilibrium and spherical symmetry, 
involves modeling the 3-D density and 3-D temperature profiles and fitting the
projected quantities to the corresponding data set.
In the projection we take into account the instrument responses 
and the effect of the  ``spectroscopic-like'' temperatures \citep{MRMT}.
The radius and mass at an overdensity of 500, $R_{500}$ and $M_{500}$
respectively, are calculated directly from the estimated cluster total
mass profile, and are listed in Table~\ref{tab-groups_dyn}. 
Note that because our clusters are
nearby, low surface brightness systems, most of the X--ray data do not
extend to $R_{500}$; in Table~\ref{tab-groups_dyn} we list the maximum
radius to which X-ray emission is detected, in units of $R_{500}$.  For
most clusters, this is $\gtrsim 70$ per cent; but clusters 7 and 15
require significant extrapolation to obtain $M_{500}$.  
The other estimates of mass and radius are included in this table, as
well.

Five of the clusters are undetected in X-rays and, for these, we
estimate an upper limit to the bolometric X--ray luminosity of about $L_{\rm X,
  bol}<3\times 10^{42}$ergs~s$^{-1}$; although the precise limit
depends on the unknown system temperature and metallicity, this limit is appropriate if
$kT>0.2$keV (Paper~II).  This is a conservative limit, corresponding to 1000 counts within
the field of view.  If the clusters were twice as large as this
field, we would still expect $\sim 250$ counts; this would be
measurable, even in the {\it ROSAT} data which is all that exists for
groups 16 and 18.  We will therefore adopt this upper limit throughout the
paper.   Corresponding upper limits on $R_{500}$, $M_{500}$ and $M_{\rm
  gas}$ are derived by fitting a logarithmic relation between these
quantities and $L_{\rm bol}$, for the 13 detected systems, and
extrapolating to the upper limit on luminosity.  This yields
$R_{500}<0.4$ Mpc, $M_{500}<1.8\times10^{13}M_\odot$, and $M_{\rm
  gas}<1.3\times10^{12}M_\odot$.  However, note that these ``limits''
on mass and radius assume that the undetected clusters lie on the same
scaling relations as brighter clusters; if they are in fact relatively deficient in
X--ray gas this is unlikely to be the case.

The X-ray image of group 15 shows a clear elongation to the north, and
is likely a double system.
This is a very low surface brightness system, from which it is
difficult to extract a reliable mass profile and hence to estimate
$R_{500}$.  The luminosity and temperature used here correspond only
to the main part of the cluster, excluding the northern elongation.  We
have applied the Anderson-Darling test, as described in \citet{Hou09}
to all groups, to test for non-Gaussianity in the velocity distribution
 This group is the only one that fails the test, with a 99.7 per cent
confident detection of non-Gaussianity.  
The spatial distribution of the galaxies is
extended in the same direction as the X-rays, but this elongation is not clearly
correlated with the velocity offsets.
Thus, in the
near-infrared data we will not attempt to separate the substructure component.

\subsection{Near-infrared data}
Of the 18 clusters in our sample,
good-quality near--infrared imaging was obtained for 11, over a period of
several years at CTIO, AAT and CFHT.  For the remaining clusters, the
only NIR data available is from 2MASS.  Table~\ref{tab-groups} lists
the basic properties of our clusters and gives the telescope and image
size of the NIR observations.  For the 2MASS clusters, the image size
just refers to the area used for our analysis --- of course the imaging
is all-sky.  

\subsubsection{CTIO}
Near-infrared data were obtained with the ISPI instrument on CTIO,
during the nights of May 8-10, 2004.  
Seeing was $\sim$1\arcsec\ in K and two of the
nights were fairly cloudy; none were photometric.  The instrument has a
$10.25\times10.25$ arcmin field of view, with 0.3 arcsec pixels.  A $3\times3$
mosaic spanning $\sim 30$ arcmin was obtained for eight
clusters, but only six were of good enough quality to include in the present
analysis.  
Data were reduced in IRAF using the {\sc cirred}
package, following the reduction algorithm described by
\citet{APDDP},  including bias
subtraction, flat fielding and sky subtraction.  The data were
calibrated by comparing the brightest galaxies with 2MASS, typically
involving a few objects per field, resulting in a zeropoint accuracy of
about $\pm 0.2$.  The mosaic was combined with the
Terapix software {\sc swarp}.

\subsubsection{AAT}
In the fall semester of 2004 we obtained data with the IRIS2
\citep{IRIS2} instrument on the AAT.
The field of view of this camera is
7.7\arcmin$\times$7.7\arcmin\ with 0.4486\arcsec\ pixels.  Six
clusters were observed, by default with a 3$\times$3 mosaic,
with a resulting coverage of 23\arcmin.  We exclude XMM5 from this
analysis, for which only 
the central position was observed.   Cluster 18 was observed in poor
conditions and several of the pointings were repeated.  The final mosaic
was made with only the best images at each position.

The data were reduced following
the same standard procedures as for the ISPI data above, but using {\sc
  starlink} software.  The astrometric  
calibration was done using the very helpful {\sc astrometry.net} software
\citep{lang09,astrometrydotnet}, and the mosaics were combined with
{\sc swarp}.  Photometric calibration was done by comparison of
bright galaxies with 2MASS, with similar accuracy to our ISPI data. 

\subsubsection{CFHT}
Four clusters were observed with WIRCAM on the CFHT during the 06A
semester.
This is a much larger
instrument and each cluster was observed with a single pointing, covering a
20 arcmin field with 0.3 arcsec pixels.  Data were reduced by Terapix in
Aug 2008.  The photometric calibration is done by comparison with
2MASS; given the larger field the zeropoint is more
precise in these fields than in our ISPI and IRIS2 data, typically
better than $\sim 0.1$ mag {\it rms}.

\subsubsection{2MASS}
For the remaining seven clusters in our sample, we do not have deep
NIR imaging so we use the public 2MASS catalogues.  These however are much
shallower than our data, with a limiting magnitude of $K<13.7$.  It is
known that the limiting surface brightness of these data is
particularly problematic, and the flux from the brightest cD galaxies
may be significantly underestimated \citep[e.g.][]{Kochanek-KLF,LM04}.
We make no explicit correction for this, and simply note that the total
luminosities for these clusters may be underestimated.

\section{Analysis}\label{sec-analysis}
\begin{figure}
\leavevmode \epsfysize=8cm \epsfbox{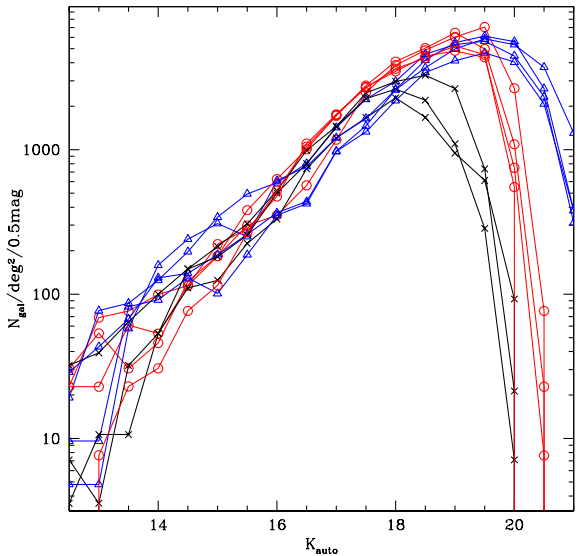} 
\caption{The number of galaxies per square degree and per 0.5 magnitude
  bin is shown for each of the 11 clusters with deep NIR observations.  The black
  lines and crosses correspond to the CTIO data; red lines and circles
  represent the AAT data; and the blue lines and triangles represent
  CFHT data.\label{fig-nc} }
\end{figure}
\begin{figure}
\leavevmode \epsfysize=8cm \epsfbox{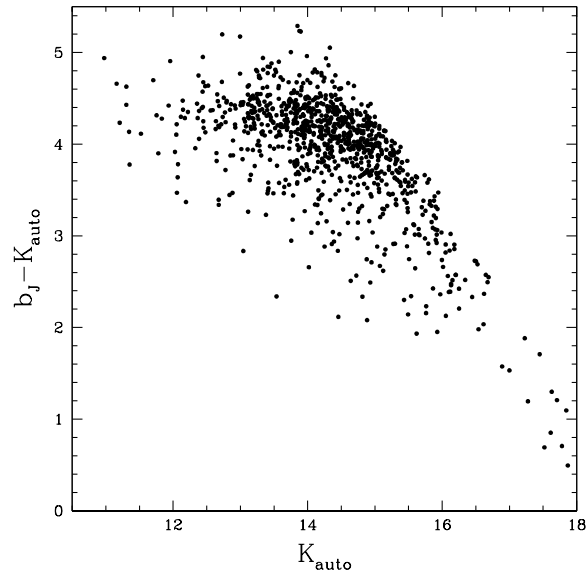} 
\caption{The colour-magnitude relation for observed galaxies with
  redshifts within 1500 km/s of each cluster.  Redshifts from the 2dFGRS
  were obtained only for galaxies with $b_J<19$; thus our sample
  becomes incomplete for red galaxies with $K>14.5$.\label{fig-bkk} }
\end{figure}
\begin{figure}
\leavevmode \epsfysize=8cm \epsfbox{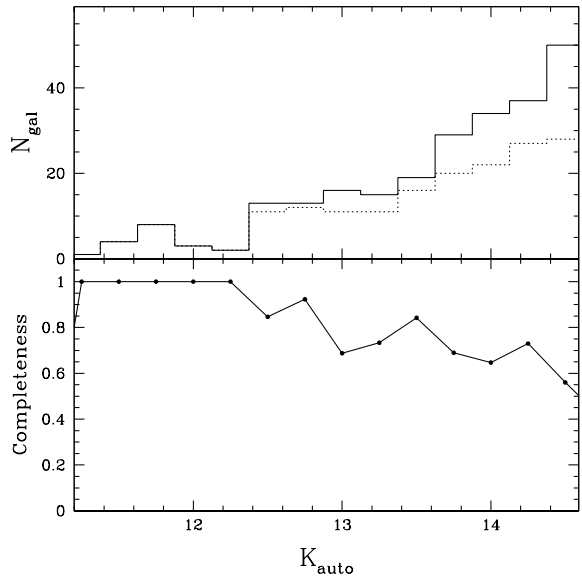} 
\caption{Top panel: The solid line shows the $K$ magnitude distribution
  of all detected galaxies in our images, while the dotted line shows
  the distribution of galaxies with redshifts from the 2dFGRS.  Bottom
  panel: The connected points represent the sampling fraction, which is
  the fraction of $K$--detected galaxies with redshift, as a function
  of magnitude.  Shown here is the average selection function; in
  practice we compute this separately for each cluster.  Only galaxies
  within $600$ kpc of the cluster centre are included.  \label{fig-zweight} }
\end{figure}
In this section we discuss the data analysis from the eleven clusters
observed at CTIO, AAT or CFHT.
Fluxes were measured from the reduced data using {\sc SExtractor} v2.5.0 \citep{sextractor}.  We
use {\sc MAG$\_$AUTO} as the best estimate of the total magnitude.
However, star-galaxy separation is done by comparing 4\arcsec\ aperture
magnitudes with the Kron radius.  The stellar locus is easily identified,
separately for each image, and we use this rather than {\sc
  CLASS$\_$STAR}.  This is only relevant for our analysis when calculating the weights to
correct for the spectroscopic sampling rate, discussed below.

We show the number of galaxies per square degree in each cluster in
Figure~\ref{fig-nc}.  The turnover in the counts gives an indication of
the depth of the observations.  The CTIO data are generally
shallowest, with a limit of $K\sim 18.0$, while the CFHT data are deepest.

Our NIR photometry is matched to the original 2dF imaging catalogues,
by searching for the closest match within a 3\arcsec\ aperture.  The
2dF catalogues are blue--selected, with $b_J<19$, and this limits our
sample size.  We show a colour-magnitude
diagram of our data in Figure~\ref{fig-bkk}, restricted to
galaxies within 1500 km/s of each cluster.  The red sequence of galaxies
typical of dense environments is evident,
and the blue selection means our sample becomes incomplete for
$K>14.5$.   This is $\sim 3$ magnitudes brighter than our $K$ detection
limit, so there is no additional incompleteness in the NIR imaging.  

Brighter than $K=14.5$ we calculate a $K-$dependent sampling
completeness for each cluster, as shown
in Figure~\ref{fig-zweight}.  This is the ratio of the number of
galaxies with a redshift to the total number detected in the NIR,
considering only galaxies within $600$kpc of each
cluster centre.  In general, the
completeness is close to 80 per cent over most of the magnitude range
we're interested in, falling to 50 per cent at $K=14.5$.  We weight all
galaxies by the inverse of this number for all subsequent analysis.
In practice the weight is calculated separately for each cluster based on
a smooth fit to the completeness function in the relevant field.

We use the mean redshift of each cluster to calculate absolute magnitudes
for all members, using a cosmology of $\Omega_m=0.3$,
$\Omega_\Lambda=0.7$, and $h=0.7$.  The most distant cluster in our sample is
at $z=0.096$, which corresponds to a luminosity distance of $442.0$
Mpc.  We correct for Galactic extinction using the \citet{SFD} dust
maps (though this is entirely negligible), and apply a k-correction of $k(z)=-6\log(1+z)$, following
\citet{Kochanek-KLF}.  For our most distant cluster, therefore, we are complete in luminosity for $M_K<-23.5$ .

\begin{figure}
\leavevmode \epsfysize=8cm \epsfbox{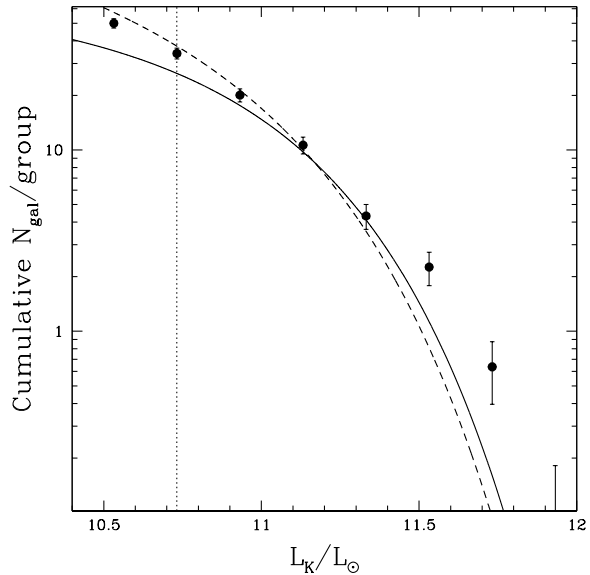} 
\caption{The cumulative luminosity function for our eleven clusters with
  deep NIR imaging.
  This includes only galaxies within 1500km/s of the cluster redshift,
  and within $R_{rms}$ of the centre.  The dotted line indicates the
  luminosity limit corresponding to $M_K>-23.5$; at luminosities fainter than this our sample is incomplete due to the 
  blue selection of the 2dFGRS.  Two Schechter functions are shown to
  guide the eye; they are not fit to the data.  Both have
  $M_K^\ast=-24.3$, taken from \citep{IRLF}.  The solid line has a
  shallow faint-end slope of $\alpha=-0.5$, while the dashed line shows
  $\alpha=-1.0$.  \label{fig-totallf} }
\end{figure}

First, we present the cumulative luminosity function from the
combination of all eleven clusters with follow-up NIR data.  We include all
galaxies with redshifts within 1500 km/s of the cluster redshift, and
within a distance $R_{\rm rms}$ from the centre\footnote{For clusters 1,
  2, and 5 the maximum radial coverage is defined by the extent of the IRIS2
  imaging.}.  We
show the weighted number of
galaxies brighter than a given $M_k$ absolute magnitude, per cluster in
Figure ~\ref{fig-totallf}.  Plotted for comparison are Schechter
functions with $M_K^\ast=-24.3$ \citep{IRLF} and $\alpha=-0.5$ (solid)
or $\alpha=-1.0$ (dashed).  These are
not fit to the data, but are meant only to guide the eye.  The
presence of central bright galaxies in most of our clusters leads to an
excess relative to the Schechter function, at the bright end, as is
seen in more massive clusters \citep[e.g.][]{Popesso-II}.  Recall that, for $M_K>-23.5$
our sample is incomplete, and this limit is indicated by the dotted line.  

\section{Results}\label{sec-results}

\begin{figure}
\leavevmode \epsfysize=8cm \epsfbox{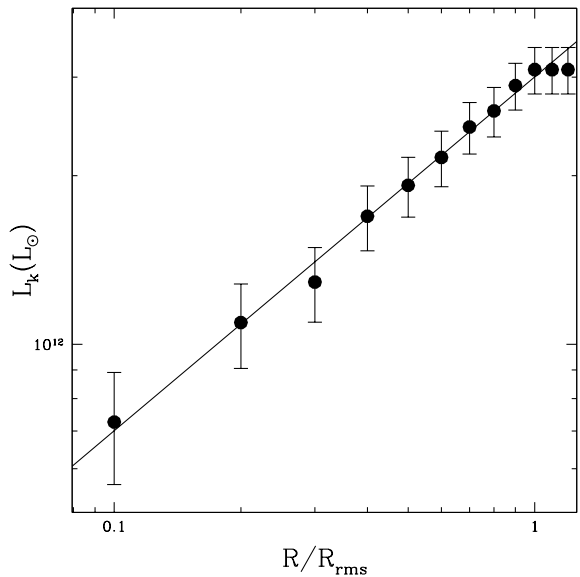} 
\caption{The cumulative $K$-band luminosity of our clusters, as a
  function of distance from the centre.  This excludes clusters with only
  2MASS photometry, or for which our NIR coverage does not extend to
  $R_{\rm rms}$.  Error bars show estimated $1\sigma$ uncertainties
  arising from Poisson fluctuations in the number of galaxies
  contributing to each bin.  The solid line is a fit to the data with
  $R/R_{\rm rms}<1.0$, and has a slope of
  $0.63$.  This relation is used to calculate how the uncertainty on
  $L_K$ depends on our uncertainty on $R_{\rm rms}$, and to correct the
  total luminosity of clusters for which our NIR coverage does not extend
  to the radius of interest.  \label{fig-dLkdr} }
\end{figure}\begin{figure}
\leavevmode \epsfysize=8cm \epsfbox{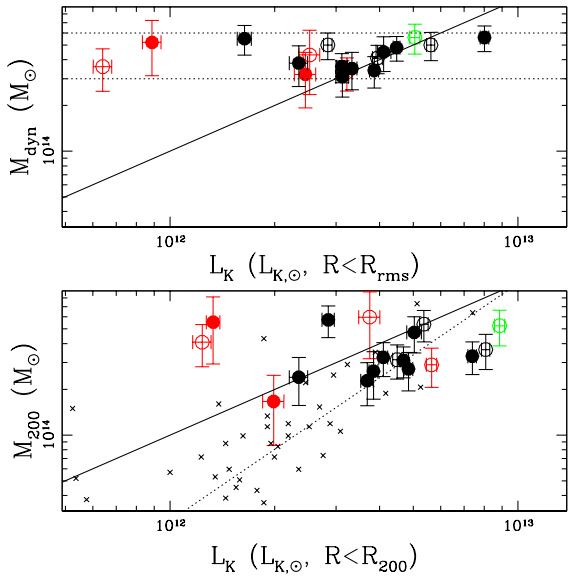} 
\caption{{\it Top panel: } The dynamical mass of our clusters is shown as a function of
  their total $K$ luminosity, integrated out to
  $R_{\rm rms}$. The horizontal, {\it dashed lines} indicate the
  selection limits imposed on the sample, while the
  solid line represents $M_{\rm dyn}/L_K=100$.    Open circles
  represent clusters with only 2MASS imaging.   The {\it green} point
corresponds to the group 15, an outlier on the $L_K-L_X$ relation  that will be identified on Figure~\ref{fig-LkLx}, while the
  {\it red} points are undetected in X-rays.  
{\it Bottom panel: }The same, but for $M_{200}$ and the $K$ luminosity
within $R_{200}$, where these quantities are measured from the velocity
dispersion alone, following \citet{RBGMR}, whose data are shown as  
the crosses
  without error bars.  The solid line represents $M_{\rm 200}/L_K=100$, while
  the {\it dotted line} represents $L_K\propto M^{0.64}$ as found by \citet{RBGMR}.  \label{fig-MdLk} }
\end{figure}

We now proceed to calculate the total $K$ luminosity of each cluster, $L_K$.  We
simply sum the luminosity of all galaxies 
within 1500 km/s and $R_{\rm rms}$ of the cluster centre, brighter than the
$K=14.5$ limit.  To correct for galaxies below this limit we model the
luminosity function as a Schechter function with $M_k^\ast=-24.3$ and
$\alpha=-1.0$.  As our data reach at least 0.6 mag fainter than
$M_k^\ast$ for all clusters, this correction is always less than 20\%.

We also include the seven clusters without deeper NIR data in our
analysis; these are shown as open symbols on the following Figures.  For these clusters, the 2MASS limiting magnitude is much brighter,
$K<13.7$, and for three of them the corresponding correction for
fainter galaxies is larger than a factor of two.  

We do not attempt to measure or correct for intracluster light.  This
remains an important uncertainty in all such work, with some claims
that a large fraction of the stars in galaxy groups are found in this
component \citep{GZZ1,McGee-ICL}.  For clusters in the mass range of
our sample, however, 
we expect the intracluster light contribution to be less than about 20
per cent \citep[e.g.][]{Zibetti}.
\begin{table*}
\begin{tabular}{lllllll}
Id & N                  & $L_K$($r<R_{\rm rms}$)& N                  &  $L_K$($r<R_{200}$)& N                  &  $L_K$($r<R_{500}$)\\ 
   &  ($r<R_{\rm rms}$) & ($10^{12}L_\odot$)      & ($r<R_{\rm 200}$)  & ($10^{12}L_\odot$)& ($r<R_{\rm 500}$)  & ($10^{12}L_\odot$)\\
\hline
1 & 28 & $3.12\pm0.12$ & 29 & $4.10\pm0.12$ & 29 & $4.07\pm0.12$\\
2 & 31 & $4.48\pm0.10$ & 31 & $4.69\pm0.10$ & 31 & $3.93\pm0.10$\\
3 & 15 & $2.84\pm0.08$ & 31 & $5.37\pm0.15$ & 11 & $2.25\pm0.06$\\
4 & 40 & $8.01\pm0.23$ & 39 & $7.39\pm0.21$ & 37 & $6.99\pm0.20$\\
5 & 23 & $3.33\pm0.11$ & 23 & $3.84\pm0.11$ & 21 & $2.43\pm0.10$\\
6 & 18 & $2.34\pm0.14$ & 18 & $2.34\pm0.14$ & 14 & $1.94\pm0.12$\\
7 & 27 & $3.95\pm0.13$ & 28 & $4.49\pm0.13$ & 27 & $3.95\pm0.13$\\
8 & 15 & $4.11\pm0.20$ & 20 & $5.03\pm0.25$ & 14 & $3.95\pm0.19$\\
9 & 17 & $5.61\pm0.15$ & 24 & $8.07\pm0.22$ & 7 & $1.95\pm0.05$\\
10 & 20 & $3.13\pm0.13$ & 25 & $3.69\pm0.15$ & 17 & $2.65\pm0.11$\\
11 & 14 & $3.22\pm0.11$ & 22 & $5.64\pm0.19$ & 5 & $1.09\pm0.04$\\
12 & 24 & $3.86\pm0.11$ & 31 & $4.84\pm0.14$ & 16 & $2.70\pm0.08$\\
13 & 3 & $0.64\pm0.04$ & 6 & $1.24\pm0.08$ & 2 & $0.47\pm0.03$\\
14 & 14 & $1.64\pm0.08$ & 21 & $2.85\pm0.11$ & 9 & $1.22\pm0.06$\\
15 & 13 & $5.05\pm0.17$ & 21 & $8.85\pm0.30$ & 9 & $3.71\pm0.13$\\
16 & 13 & $2.45\pm0.17$ & 10 & $1.98\pm0.14$ & 3 & $0.97\pm0.07$\\
17 & 4 & $2.52\pm0.18$ & 6 & $3.75\pm0.26$ & 3 & $1.96\pm0.14$\\
18 & 4 & $0.89\pm0.05$ & 5 & $1.33\pm0.06$ & 2 & $0.66\pm0.04$\\

\hline
\end{tabular}
\caption{Derived properties of the stellar population for all clusters
  in our sample.  Throughout the paper, stellar masses are calculated
  using a universal mass--to--light ratio in the $K-$ band,
  $\gamma=0.7$. \label{tab-groups2}}
\end{table*}

The statistical uncertainty on $L_K$ is dominated by the statistical
uncertainty on $R_{\rm rms}$, since that quantity determines the radius within
which the luminosity is integrated\footnote{Poisson-type uncertainties on the number of cluster members $N$ are
not really appropriate in this context, as for a specific cluster $N$ is
a fixed number without uncertainty.  Statistical uncertainties on the photometry, while of
order 5\%, are negligible because they are reduced by $\sqrt{N}$ when
applied to the total luminosity.}.  In Figure~\ref{fig-dLkdr} we show
the cumulative luminosity as a function of $r/R_{\rm rms}$,
only including clusters observed out
to $R_{\rm rms}$, with AAT, CTIO or CFHT. The best-fit
line to the data where $R/R_{\rm rms}<1$ has a slope of $0.63$.    Thus, the statistical
uncertainty on $L_K$ is only $\sim 0.63\Delta R/R_{\rm rms}$ which,
given the typical 10 per cent uncertainty on $R_{\rm rms}$, corresponds
to a $\sim 6$ per
cent uncertainty on $L_K$.  In contrast, the typical uncertainty on $M_{200}$ is 20--40 per cent, as it is proportional to $\sigma^3$
(see \S~\ref{sec-obs}).
We use this relation for $L_K(R)$ to correct the total luminosity of those clusters
for which NIR coverage only extends out to $r<R_{\rm rms}$.

The total $K$ luminosity of each cluster, integrated to either $R_{\rm
  rms}$, $R_{200}$ or $R_{500}$, is given in Table~\ref{tab-groups2}.  We also
show the total number of galaxies with redshifts and NIR data within
each radius.

\subsection{Correlation between stellar luminosity and dynamical mass}
We have shown above that $L_K$ can easily be measured with a precision about
five times better than that of
$M_{\rm dyn}$, a point that has been noted by others
\citep[e.g.][]{Popesso-III}.  This makes it a very useful indicator of
system mass, although of course it is tracking a fundamentally
different quantity than $M_{\rm dyn}$.  In 
the top panel of Figure~\ref{fig-MdLk} we show the correlation between
these two quantities.  The clusters were selected to span a factor of
only three
in $M_{\rm dyn}$, but they show a factor $\sim 10$ spread in $L_K$.
Note that the apparent lack of correlation is likely a consequence of
limited dynamic range in $M_{\rm dyn}$, together with significant
scatter between $M_{\rm dyn}$ and $L_K$. 
Most of the clusters are consistent with $M_{\rm dyn}/L_K=100$, shown
as the solid line.
The green and red points indicate X-ray underluminous
systems, which will be discussed below.  Note these most of these have $M_{\rm
  dyn}/L_K$ ratios in good agreement with the rest of the sample.

To compare with data from  \citet{RBGMR} we compute $R_{200}$ and
$M_{200}$ in precisely the same way they do, using equations \ref{eqn-r200} and \ref{eqn-m200}.  
We measure $L_K$ also within $R_{200}$, and
show the correlation with $M_{200}$ in the bottom panel of
Figure~\ref{fig-MdLk}. The data of \citet{RBGMR} are
shown as the small crosses.
The solid line represents a mass-to-light ratio of
$M_{\rm 200}/L_K(r<R_{200})=100$, while the dotted line shows the relation
found by \citet{RBGMR}, $L_K\propto M^{0.64}$.  
Most of the clusters are consistent with the data of \citet{RBGMR},
though there are clear outliers.  Most notable are the two groups (13
and 18) that have stellar masses well below that expected from their
dynamical mass.  The fact that they are undetected in X-rays suggests
that their dynamical mass is significantly overestimated; however, we
will show (Fig \ref{fig-LkLx}) that the lack of X-ray emission is still
surprising given their total stellar mass.

\subsection{X-ray properties}
\begin{figure}
\leavevmode \epsfysize=8cm \epsfbox{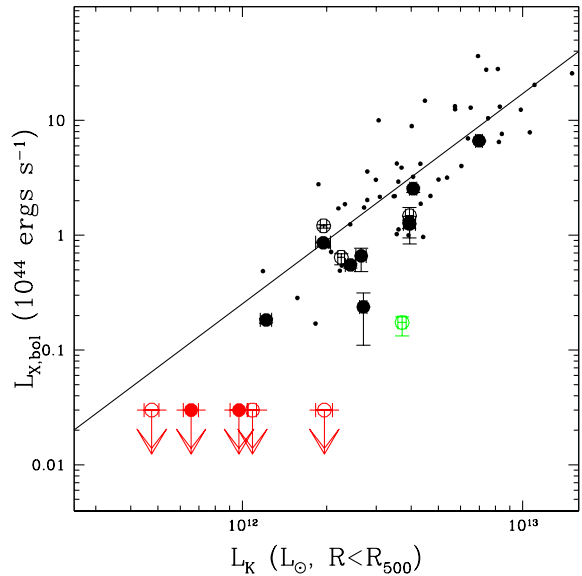} 
\caption{The bolometric X--ray luminosity of our clusters are compared
  with $L_K$; both quantities are computed within
  $R_{500}$ as determined from the X--rays.  The small, filled circles
  are the clusters of \citet{LM04}, with X-ray luminosities taken from
  \citet{HIFLUGCS}, and the solid line is a linear fit to these data. 
The X-ray undetected clusters (red points)
  are represented as upper limits, with $L_K$ measured within $R_{\rm
    500}\sim 0.4$ Mpc.  The cluster marked in green is detected in X-ray, but is a clear outlier on this plot, with about five times more near-infrared luminosity than expected.  \label{fig-LkLx} }
\end{figure}
In Figure~\ref{fig-LkLx} we show the bolometric X--ray luminosity of
each cluster as a
function of its total $K-$band luminosity.  Both quantities are computed
within $R_{500}$ as determined from the X-ray emission (Paper~II).
Throughout the paper, both the bolometric luminosity $L_X$ and X-ray
temperature include contribution from any cool-core.  This maximizes
the scatter and allows us to test for any correlation between cool-core
emission and optical properties of the clusters.
We compare our measurements
with the data of \citet{LM04}, with X-ray luminosities
(also uncorrected for cool cores) obtained
from \citet{HIFLUGCS}; the solid line represents a least-squares fit to
these data.  There is considerable scatter among our
clusters, but most of the X--ray detected clusters are 
consistent with those of \citet{HIFLUGCS}.  There is some indication that our clusters lie toward
the lower edge of the distribution.  This is likely due to the
well-known Malmquist bias that affects X--ray selected samples
\citep[e.g.][]{Ikebe,PCAB,Vik+09,Rykoff2}.  We therefore expect our
dynamically-selected cluster sample to have a representative
distribution of $L_X$.  
\begin{figure}
\leavevmode \epsfysize=8cm \epsfbox{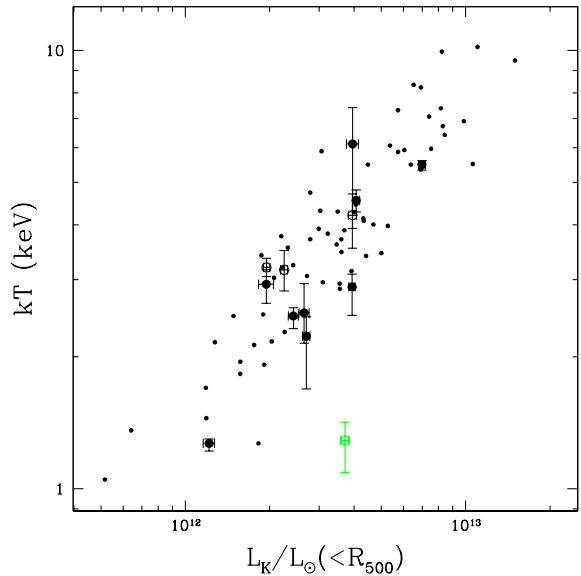} 
\caption{The X--ray temperature of our 13 detected clusters are compared
  with $L_K$ as measured within
  $R_{500}$ .  The small, filled circles
  are the clusters of \citet{LM04}, with X-ray temperatures taken from
  \citet{Hornerthesis}.  The green symbol corresponds to group 15,
which is also an outlier in figure~\ref{fig-LkLx}.\label{fig-LkT} }
\end{figure}

One of the systems (group 15,
in green) appears to have significantly low $L_X$ for its stellar
luminosity.  Recall from \S~\ref{sec-xray} that group 15 shows
substructure in the X-ray and a significantly non-Gaussian velocity
distribution.  The luminosity and temperature used here
correspond only to the dominant component.   Treating the whole system
as one results in a temperature increase of $\sim 10$ per cent, and a
$L_X$ increase of $\sim 73$\%.  This does not change the status of this
system as an outlier on this plot.

The limits on the five undetected clusters imply that they
are underluminous in X--ray by up to a factor of 10.  Recall that,
for these clusters the radius $R_{500}$ is based on the
radius expected for a cluster at our $L_{\rm bol}$ detection limit.  If
the clusters are actually deficient in X--ray emitting gas, the true
value of $R_{500}$, and hence $L_K$, could be be larger.  This would
make the discrepancy worse, since the limit on $L_X$ is based on a
fixed number of counts in the field of view.

\begin{figure}
\leavevmode \epsfysize=8cm \epsfbox{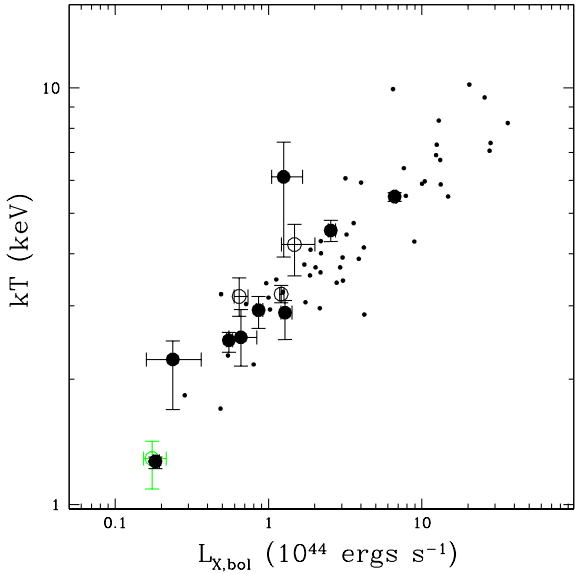} 
\caption{The correlation between X--ray luminosity and temperature for
  the 13 detected clusters, compared with the \citet{LM04} cluster
  sample.  The green point indicates group 15, an outlier on the
  $L_K-L_X$ relation. The fact that it appears consistent with the rest
  of the data shown here suggests that the offset is due to unusually
  high $L_K$.  \label{fig-LxTx} } 
\end{figure}

The total X--ray luminosity is known to be a relatively poor tracer of mass, but a good tracer of the thermodynamic history of the gas
\citep[e.g.][]{scatter}.  The X-ray temperature, while more
difficult to measure, is a much better representation of the size of
the potential.  Thus, in Figure~\ref{fig-LkT} we show the correlation
between temperature and $L_K$ for the 13 clusters in our sample with
X-ray detections.  The temperature measurements are described in Paper~II.  Again, these data are compared with data from
\citet{LM04}, with temperatures (uncorrected for cool cores) from \citet{Hornerthesis}.  Our data
generally agree very well with the relation defined by the \citet{LM04}
data, confirming that $L_K$ is a good tracer of cluster mass.
Interestingly, the detected cluster 15 (in green), that is a significant outlier on the $L_K-L_X$ relation, also lies off the $L_K-T$ relation in Figure~\ref{fig-LkT}.  This could indicate that it is a high stellar content, rather than low X--ray luminosity, that makes it unusual.  This is also seen in the $L_X-T$ relation, shown in Figure~\ref{fig-LxTx}, where our data are compared with those of 
\citet{Hornerthesis}.  Group 15 appears consistent with this relation.
We again note, however, the possible, small offset of the whole sample
toward lower $L_X$ relative to \citet{LM04}, indicative of Malmquist
bias in X--ray flux-limited samples (Paper II).

\begin{figure}
\leavevmode \epsfysize=8cm \epsfbox{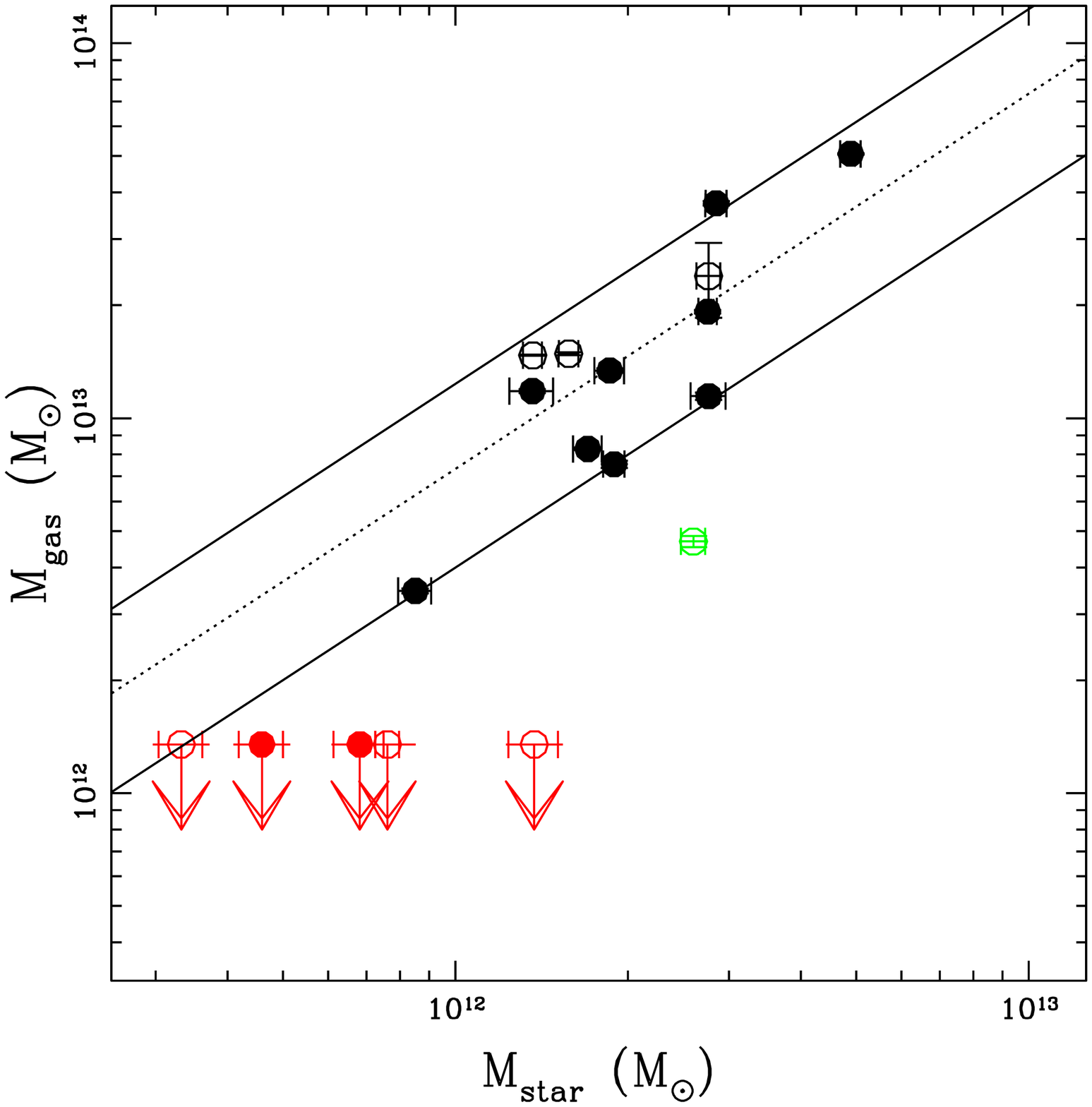} 
\caption{The total gas mass within $R_{500}$, as measured from the X--ray emission, is
  compared with the total stellar mass within the same radius, computed from $L_{K}$ assuming
  $\gamma=0.7M_\odot/L_\odot$.  Lines represent different values of $f=M_{\rm star}/\left(M_{\rm
    star}+M_{\rm gas}\right)$ for comparison.  The dotted line shows
$f=0.12$, while the two solid lines show $f=0.075$
(top line) and $f_{\rm star}=0.2$ (bottom).  
The X-ray undetected clusters
  are represented as red upper limits, with $L_K$ measured within $R_{\rm
    500}\sim 0.4$ Mpc.   The green symbol represents group 15, the outlier on Figure~\ref{fig-LkLx}.
  \label{fig-fstar} }
\end{figure}

\subsection{Stellar mass}
Finally, we convert the $K-$luminosity into a stellar mass, assuming a
stellar mass-to-light ratio $\gamma=0.7M_\odot/L_\odot$ \citep{LM04}.
In Figure~\ref{fig-fstar} we compare this with the gas mass within
$R_{\rm 500}$, as measured from the X--ray observations.  The lines represent different values of the baryon
fraction found in stars, $f=M_{\rm star}/\left(M_{\rm
    star}+M_{\rm gas}\right)$, of 0.075, 0.12 and 0.2.  Our data show
considerable scatter, but on average our detected clusters are consistent
with $\sim 12$ per cent 
of their baryons in stars (neglecting intracluster light), as typically
found for more massive clusters \citep[e.g.][]{baryons}.   For the five
undetected clusters, we show the estimated upper limit of $M_{\rm
  gas}<1.3\times10^{13}M_\odot$, and compute the stellar mass within
the radius $R_{500}\sim 0.4$ Mpc that would be consistent with an otherwise
normal cluster at our X--ray detection limit.   These limits
imply that $>20$ per cent of the expected baryons are in the form of
stars.   If $R_{500}$ is larger than we have estimated, for example if
their dynamical masses are correct, then $M_{\rm star}$ could be
substantially larger.  Thus, our limit on the stellar fraction here is
robust.  These are clearly very different 
systems from the rest of the sample, with little or no associated X--ray emitting gas.
\begin{figure}
\leavevmode \epsfysize=8cm \epsfbox{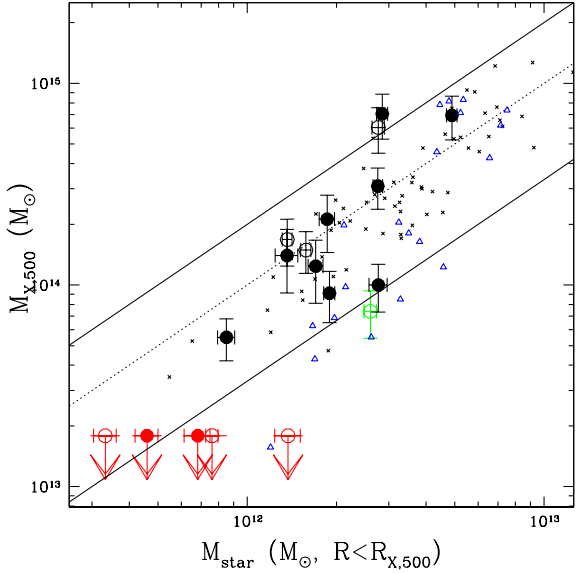} 
\caption{The total cluster mass within $R_{500}$, as measured from the X--ray emission, is
  compared with the total stellar mass within the same radius, computed from $L_{K}$ assuming
  $\gamma=0.7M_\odot/L_\odot$.  Lines represent different values of $f=M_{\rm star}/M_{\rm 500}$ for comparison.  The dotted line shows
$f=0.01$, while the two solid lines show $f=0.005$
(top line) and $f_{\rm star}=0.03$ (bottom).  
The X-ray undetected clusters
  are represented as red points, and the green symbol represents group
  15, the outlier on Figure~\ref{fig-LkLx}.  Small crosses are data from
  \citet{LM04}, and blue triangles are clusters from \citet{GZZ},
  excluding the intracluster light component, as presented in
  \citet{BMBE}.  With the exception of the single outlier, all our
detected clusters have total stellar fractions less
  than 3 per cent, with an average of about 1 per cent.   These
  stellar fractions are consistent with those of
  \citet{LM04}, but lower than found by \citet{GZZ} in their
  lowest-mass systems.  
  \label{fig-fstar2a} } 
\end{figure}

\begin{figure}
\leavevmode \epsfysize=8cm \epsfbox{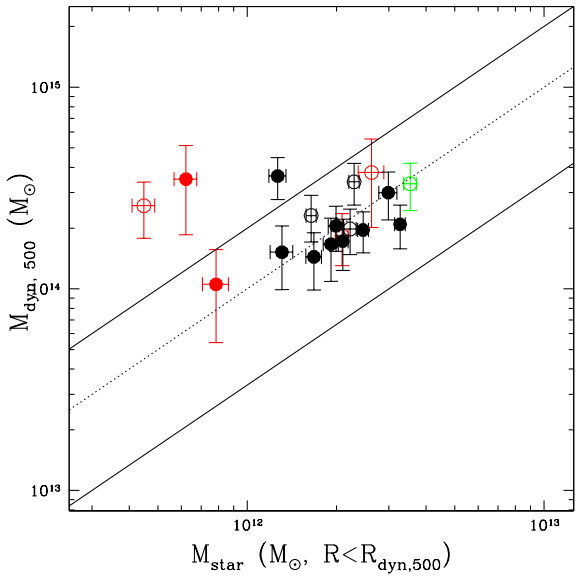} 
\caption{The same as Figure~\ref{fig-fstar2a}, but where $R_{500}$ and $M_{500}$ are now
estimated from the galaxy dynamics, for all clusters in our sample.  
  \label{fig-fstar2b} } 
\end{figure}

We next compare the stellar mass with the total
mass $M_{500}$, as estimated from either the X--ray mass profiles
(Figure~\ref{fig-fstar2a}), or the galaxy dynamics (Figure~\ref{fig-fstar2b}).  In the latter case,  we 
use $R_{500}=\sqrt{200/500}R_{200}$, and $M_{500}=\sqrt{200/500}M_{200}$. 
The lines in this figure represent total stellar fractions
of $M_{\rm stars}/M_{\rm 500}=0.005$, $0.01$ and $0.03$.  The data are
compared with \citet{LM04} and \citet{GZZ}.  For the latter, we exclude
the intracluster light component, as neither our data nor
that of \citet{LM04} includes this.  We omit the two clusters in
the \citet{GZZ} sample that are strongly affected by line-of-sight
projections; the two lowest-mass clusters remaining in their 
sample have the most uncertain $R_{200}$ and \citet{GZZ} exclude them
for some of their analysis, though we retain them here for the
sake of completeness.

In general the agreement
with the \citet{LM04} data is good, but our lowest-mass clusters
($M_{\rm star}<4\times10^{12}M_\odot$) have
systematically lower stellar fractions than similar systems from
\citet{GZZ}.  In particular, 
none of the X--ray detected  clusters
have total stellar fractions greater than about 3 per cent, and most
are closer to 1 per cent.   
Our data are in good agreement
with the theoretical prejudice of \citet{BMBE}, that the stellar
fraction in groups and clusters cannot be much larger than observed in
the most massive systems. 

Some of the five clusters that are undetected in X--rays
appear to have unusually high stellar fractions; however recall that
the radius $R_{500}$ within which the stellar light is integrated is
not directly measured for these systems. In Figure~\ref{fig-fstar2b} we use 
dynamical measures of $M_{500}$ and $R_{500}$, which we can measure
uniformly for
the whole sample.  
Here we see that all our clusters are
remarkably uniform, with a total stellar fraction close to $\sim 1$ per
cent in most cases.  

Finally, we note that,
for the most part, the open symbols on
Figures~\ref{fig-MdLk}--\ref{fig-fstar}, which indicate the clusters
with only 2MASS imaging, do not appear distinct from the 
other points on the plots.  This gives confidence that our results are
not significantly biased by the lack of deeper imaging in these
systems.  The obvious exception is group 15, which is an outlier on
most plots.  This, however, may be related to the fact that it is a
clear double-system, and the only cluster with a significantly
non-Gaussian velocity distribution.

\section{Discussion}\label{sec-discuss}
Of the 18 clusters observed, five are undetected in X--ray, and one
(group 15) appears to have an anomalously high stellar content given
its X--ray properties.  The five non-detections have a stellar mass
which is $\sim 1$ per cent or less of their total dynamical mass,
consistent with most other clusters.  However, the strong limits on
their X--ray luminosity imply that there is little or no associated
X--ray emitting gas.  For these systems, either  
their {\it virialised} mass is 
$M< M_{\rm dyn}$, or they are extremely 
  underluminous due to excess heating or cooling.   
\begin{figure}
\leavevmode \epsfysize=8cm \epsfbox{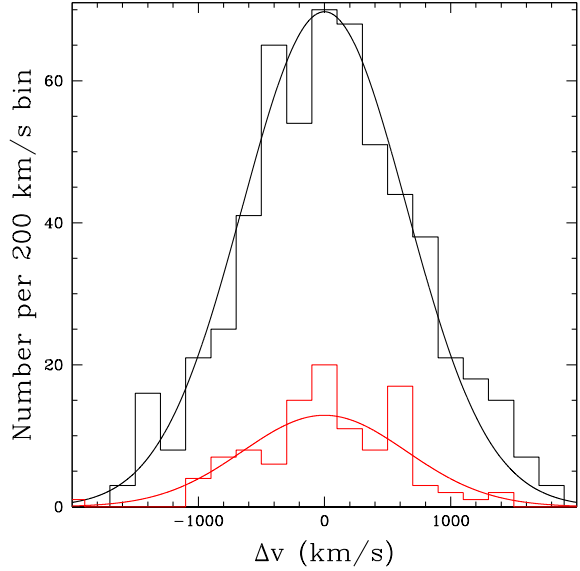} 
\caption{The velocity distribution for our cluster sample, considering
  galaxies within $1$ Mpc of the recomputed centre.  The {\it red} histogram
  corresponds to the five undetected systems.  The rest of the
  clusters are included in the {\it black histogram}.  The smooth lines
  are Gaussian functions with $\sigma=650$km/s, normalized to match the
  total number of objects in each histogram.  \label{fig-vdistall} }
\end{figure}

Of these, the least likely possibility is that all the gas has cooled
to form stars.  
Figure~\ref{fig-fstar2b} suggests that, if anything, the ratio of stellar
mass to dynamical mass in these clusters is {\it
  lower} than that of the normal systems.  Normal clusters have $\sim
7$ times more gas mass than stellar mass (Figure~\ref{fig-fstar}); if
all that gas formed stars we would expect to see stellar luminosities
$\sim 7$ times higher than normal clusters of the same dynamical mass,
and this is clearly not the case.

On the other hand it is difficult to rule out the possibility that the
gas in the underluminous clusters has been heated or expelled, as predicted by some models \citep[e.g.][]{Bower+08,McCarthy+10}.  In this
case of course there should be some gas present, but at low density and
low surface brightness.  The only way to definitively distinguish
between this and the following scenario is through X-ray
observations deep enough to detect either this hot, diffuse gas or the
gas associated with individual galaxies.

This leaves the possibility that the undetected clusters are 
  unrelaxed, and have not yet reached virial equilibrium
  \citep[e.g.][]{Popesso-V,D+09,XI-I}.  It is even possible that some
  of these systems are chance projections ,
  in which case
  the appropriate ``virialised mass'' to consider is just that of the
  dominant galaxy.  To shed light on this issue we now investigate
  this using the spatial and dynamic information available for each
  cluster.  Since these results are somewhat sensitive to the choice of
  centre and radius, we recompute these quantities for all the
  following analysis, including galaxies with redshifts from surveys
  other than the original 2dFGRS.  Specifically, we select all galaxies
  with redshifts, within 1.5 Mpc of the original
  centre and 1500 km/s of the original cluster redshift.  We adopt the
  central position and redshift as the geometric mean of these
  quantities, unweighted for selection or luminosity.  
  
In Figure~\ref{fig-vdistall} we show the velocity distributions for
the "normal" clusters (black line), including only galaxies within 1 Mpc of the
recomputed centre.  We compare this with the five undetected clusters, as the red line. 
Overplotted for comparison are Gaussian functions with
a velocity dispersion of $650$km/s, for comparison.  
A Kolmogorov--Smirnov test shows that the velocity distributions of
the undetected clusters are consistent with that
of the rest of the sample, with a 13 per cent probability of either being
drawn from the same parent distribution.  Thus, given the limited statistics, we find
no evidence for differences in the dynamical state of the underluminous
systems.  As described in \S~\ref{sec-xray}, we have also computed
the Anderson-Darling statistic, as described by \citet{Hou09}, for each
group.  Only group 15 shows significant non-Gaussianity in the velocity
distribution, at $>95$ per cent confidence.
\begin{figure}
\leavevmode \epsfysize=8cm \epsfbox{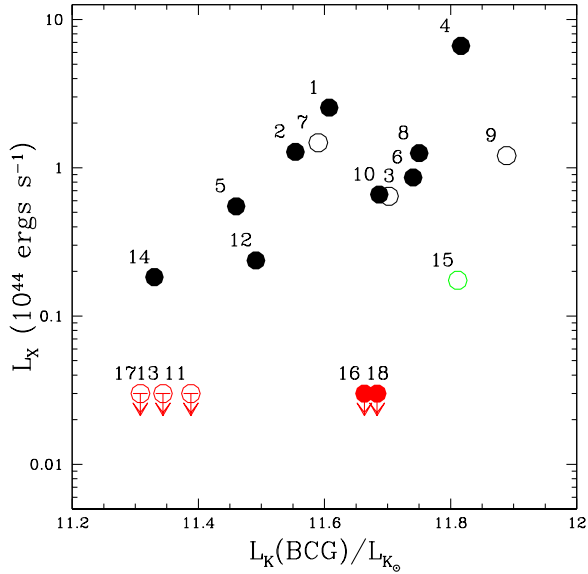} 
\caption{For each cluster, we show the cluster X--ray
  luminosity as a function of the $K$-band luminosity of the
  brightest cluster member (BCG).  Point symbols are as in other plots, with the {\it green}
  symbol for the outlier group 15, and {\it red} symbols for undetected
  clusters.  Clusters are numbered according to their entry in
  Table~\ref{tab-groups}.  Open symbols represent clusters for which
  the near-infrared imaging comes from 2MASS.  The lack of correlation
  here demonstrates that the clusters which are underluminous in X--ray
  host BCGs that are just as luminous as the rest of the sample.\label{fig-bcg_mag} }
\end{figure}  
\begin{figure}
\leavevmode \epsfysize=8cm \epsfbox{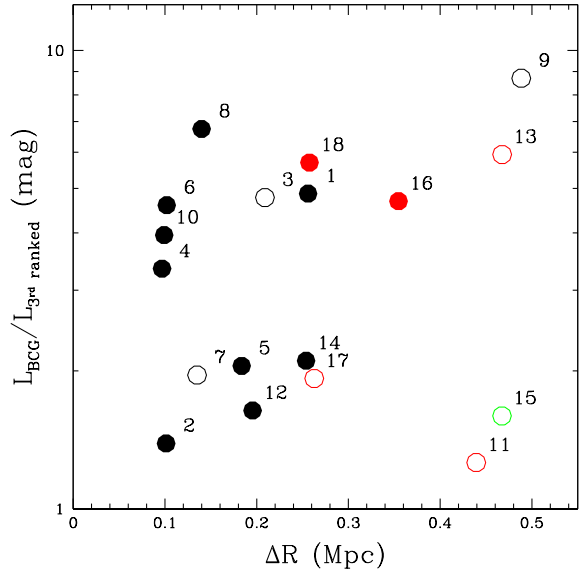} 
\caption{For each cluster, we show the luminosity ratio of the brightest
  and third-brightest galaxies (in $K$), as a function of the distance of
  the brightest galaxy from the recomputed cluster centre.  Symbols are
  as in Figure~\ref{fig-bcg_mag}. \label{fig-bcgs} }
\end{figure}

Other indications of a relaxed cluster could be the
presence of a dominant galaxy near the X-ray centre \citep[e.g.][]{D+10}, and an
approximately spherical, centrally--concentrated galaxy distribution.
 In many of the
clusters, there is an obvious dominant galaxy, large and luminous, near the centre of the
X-ray image.  
To quantify this, we have selected all cluster members
within 700 km/s of the mean redshift and 500 kpc of the recomputed
centre, and identified the most luminous (in $K$) as the BCG.
Figure~\ref{fig-bcg_mag} shows the luminosity of each of these BCGs, as
a function of the host cluster $L_X$.  
There is little
correlation here, although we note that three of the X-ray undetected
clusters have BCGs that are among the least luminous in the sample
(unfortunately, all three have only shallow 2MASS imaging, and thus the
total luminosities may be significantly underestimated). It is also interesting that group 15, which has a high stellar mass given its $L_X$, has one of the most massive BCGs in the sample.

Next we
calculated the luminosity ratio between the first- and third-ranked galaxy, $L_{K,13}$.
In Figure~\ref{fig-bcgs} we show this as a function of
$\Delta R$, which is the distance in Mpc between the brightest cluster
galaxy (BCG) and the
geometric centre of the cluster, recomputed as above.   
Interestingly, the distribution of the X-ray undetected clusters (and also the
outlier group 15) is distinct from most of the ``normal'' systems, in
the sense that their BCG is at least 250 kpc from the centre.  However,
we note that, of the 12 otherwise ``normal'' systems, only about half
have a dominant ($L_{K,13}>3$), centrally located ($\Delta R<250$ kpc) BCG.  

\begin{figure}
\leavevmode \epsfysize=8cm \epsfbox{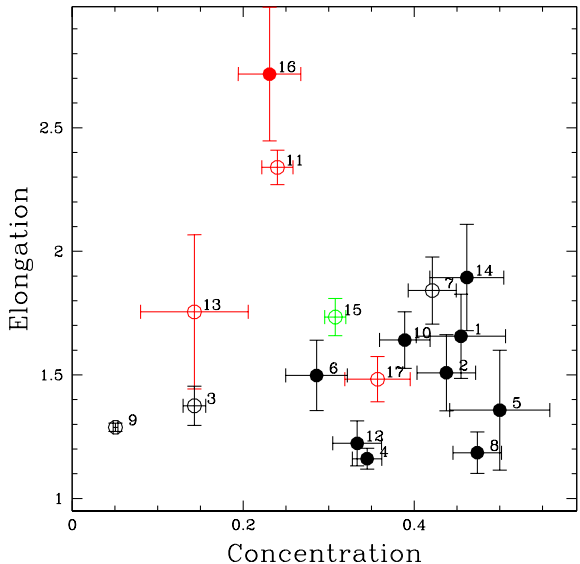} 
\caption{For each cluster (except cluster 18, which has too few members for
  this analysis), we show its elongation as a function of
  concentration.  Both quantities are computed for a subset of cluster
  members, brighter than $M_K^\ast+0.6$ and with velocities within twice
  the cluster velocity dispersion. The concentration is the fraction of
  galaxies that lie within $0.5R_{\rm rms}$, while the elongation is
  the ratio of the {\it rms} values about the directions perpendicular
  to and parallel to the least-squares fit correlation in spatial
  coordinates.  Symbols are the same as figure~\ref{fig-bcgs}.  Three
  of the undetected clusters show low values of concentration, and two
  of these are also highly elongated.
\label{fig-shapes} }
\end{figure}

We now attempt to quantify the spatial distribution of 
the most luminous galaxies, i.e. those that are at most 0.6 mag fainter than
$M_K^\ast$, so that we are equally deep in all clusters.  The
undetected cluster 18 only
has three galaxies above this limit, so we omit it from the following
analysis.  We calculate the concentration as the fraction of such
cluster members within $0.5R_{\rm rms}$.  For the elongation, we first
perform a least-squares regression analysis to find the principle axis
of each cluster on the sky; then we calculate the {\it rms}
dispersion perpendicular to and parallel to this axis.  The elongation
is the ratio of the two values, always defined as the larger divided by
the smaller so the ratio is greater than unity.  
The results are shown in Figure~\ref{fig-shapes}, where the points are
colour-coded as before.  Error bars are computed using a jackknife
resampling. Interestingly, most of the X-ray underluminous systems again appear
separated from the majority of the population, as either low-concentration
or highly elongated clusters.  Only cluster 17 lies in a region of the
plane occupied by the majority of normal clusters.  Note that cluster
16 is highly elongated; from Figure~\ref{fig-bcgs} we see that it
has a distinctly dominated galaxy, but located 350 kpc from the
centre.  This may indicate a merging or otherwise unrelaxed system.
Cluster 18, which is the only other X--ray undetected system with a
dominant, centrally-located galaxy, has too few members to measure either quantity shown
here with adequate precision.   Again, however, there are
examples of clusters (3 and 9) with normal X-ray properties and equally low concentrations.

This result needs to be approached with some
caution, as there are multiple parameters at work here, related to the
magnitude and velocity selections, the choice of centre, and the
definition of concentration.  Nonetheless, we tentatively conclude that
it seems likely dynamical age plays some role in 
the X-ray luminosity of a given cluster.  All the clusters that appear
relaxed in the optical -- with a dominant central galaxy (within 250kpc
of the cluster centre, and at least three times brighter than the
third--ranked galaxy), a
centrally concentrated ($>0.25$) galaxy population, and a spatial axis ratio of
less than two -- show normal X-ray properties\footnote{The possible
  exception is group 18, which has a dominant, near-central galaxy but
  for which we are unable to measure elongation and concentration.}.  There is no evidence for excess heating
or cooling in these systems.  However, this does not mean that X--ray
selection returns a sample of relaxed clusters; of the twelve
clusters with normal X-ray properties (i.e. excluding the outlier 15 and the
undetected clusters), seven show signs of dynamical youth.  In other words, neither optical- nor X-ray
selection returns a sample of dynamically relaxed clusters, which is
certainly not a surprise.  It is worth emphasizing though that any
signs of dynamical disturbance are fairly subtle.  With the possible
exception of group 15, none of our clusters
look like strongly merging systems either in the X--ray or optical
images, and the velocity distributions (based on few members) do not
show obvious signs of nonvirialisation.  

Thus, with the present data, we  
can plausibly argue that the undetected systems might be  
unvirialised and that this
is the origin of the low X-ray emission. On the other hand, the  
differences between the luminous and under-luminous systems are  
remarkably subtle and we cannot rule out that the difference is due to  
a substantial amount of ejected gas, as predicted by some models \citep[e.g.][]{Bower+08,McCarthy+10}.

\section{Conclusions}\label{sec-conc}
We have presented a sample of 18 clusters selected from
\citet{Eke-groups} based on their
dynamical mass alone, and followed up sixteen of them with pointed {\it Chandra} and {\it
  XMM} observations.  The clusters were selected based on optical properties alone, to lie in a narrow range of
dynamical mass $3\times10^{14}<M/M_\odot<6\times10^{14}$.  We summarize our findings as follows: 
\begin{itemize}
\item Of the 18 clusters studied,
  five are undetected in X--ray emission.  The limits on their X--ray
luminosity are significantly below the luminosity that would be
expected given their stellar mass.  The
  rest lie within the scatter defined by X--ray selected samples,
  but
  toward the low--Lx end of the scatter.  This is in good agreement with previous
  studies by e.g. \citet{BCECB}, \citet{Rykoff1},  \citet{Hicks08} and \citet{XI-I}.
\item Stars make up less than 3 per cent of the total mass in all of
  our clusters, with an average of about 1 per cent.  The fraction of detectable baryons in the form of
  stars is about 12 per cent on average, with a range of approximately
  $7-20$ per cent, though the clusters that are undetected
  in X--ray emission have implied fractions of $>20$
  per cent.  
\item The undetected clusters have velocity distributions that are not
  significantly different from the rest of the sample.  However, they
  all either lack a central, dominant galaxy, or show spatial
  distributions that are of low concentration or high elongation,
  relative to most of the ``normal'' clusters.  Similar conclusions have
  been reached by \citet{Popesso-V} and \citet{D+09}.
\end{itemize}

We conclude that the redshift--selection of low--mass clusters returns
a heterogeneous sample.  Most of those systems that are detected in
X--ray have gas properties that are consistent with X--ray selected
samples, though somewhat underluminous on average.  The main
difference is that a fraction (6/18 in our case) have surprisingly low
X--ray luminosity, given their stellar content.  It remains an open
question, whether this is because such clusters are dynamically young
(unvirialised) or because the hot gas has been expelled or rarefied due
to energetic processes associated with galaxy formation.

\section{Acknowledgments}\label{sec-akn}
We thank the referee for a thorough and helpful report, and Stefano
Andreon for pointing out a few errors in the original text.
MLB and TL would like to thank David Gilbank and Sean McGee for many useful
discussions about this work and comments on the draft.  
MLB would like to thank to Patrice Bouchet for carrying out our CTIO
observations when we were unable to be present for our run.
We thank Terapix, in particular Patrick Hudelot, for their help in
reducing CFHT WIRCAM data.  MLB would also like to thank Sean McGee for
calibrating our AAT data using {\sc astrometry.net}.  This research has
made use of the NASA/IPAC Extragalactic Database (NED) which is
operated by the Jet Propulsion Laboratory, California Institute of
Technology, under contract with the National Aeronautics and Space
Administration.  
This research is supported by an NSERC Discovery grant to MLB. PM and
HB acknowledge support by contract ASI-INAF I/023/05/0, ASI-INAF
I/088/06/0, NASA grants NNX09AP45G and NNX09AP36G.

\bibliography{ms}
\end{document}